# Making brain-machine interfaces robust to future neural variability

David Sussillo[1,5,*], Sergey D. Stavisky[2,*], Jonathan C. Kao[1,*], Stephen I Ryu[1,7], Krishna V. Shenoy[1,2,3,4,5,6]

*Equal Contribution
[1]Electrical Engineering Department
[2]Neurosciences Graduate Program
[3]Neurobiology and Bioengineering Departments
[4]Bio-X Program
[5]Stanford Neurosciences Institute
[6]Howard Hughes Medical Institute at Stanford University
Stanford University, Stanford, 94305 CA, USA.
[7]Palo Alto Medical Foundation
Palo Alto, 94301 CA, USA

Contact: shenoy@stanford.edu

## ABSTRACT

A major hurdle to clinical translation of brain-machine interfaces (BMIs) is that current decoders, which are trained from a small quantity of recent data, become ineffective when neural recording conditions subsequently change. We tested whether a decoder could be made more robust to future neural variability by training it to handle a variety of recording conditions sampled from months of previously collected data as well as synthetic training data perturbations. We developed a new multiplicative recurrent neural network BMI decoder that successfully learned a large variety of neural-to-kinematic mappings and became more robust with larger training datasets. When tested with a non-human primate preclinical BMI model, this decoder was robust under conditions that disabled a state-of-the-art Kalman filter based decoder. These results validate a new BMI strategy in which accumulated data history is effectively harnessed, and may facilitate reliable daily BMI use by reducing decoder retraining downtime.

Brain-machine interfaces (BMIs) can restore motor function and communication to people with paralysis[1,2]. Progress has been particularly strong towards enabling two-dimensional computer cursor control, which may allow versatile communications prostheses[3–5]. Cursor control performance has approached that of the native hand in recent macaque studies[6,7], but this was done under favorable laboratory conditions where neural recordings are often stable both during and across BMI sessions[8–11]. In contrast to these pre-clinical studies, one of the major challenges impeding BMI use by human clinical trial participants is the high degree of within- and across-day variability in neural recording conditions (Fig. 1a)[12–16]. We use the term "recording condition" to broadly encompass the combination of factors which together determine the relationship between observed neural activity and intended kinematics. These factors include the relative position of the electrodes and surrounding neurons (diagrammed in Fig 1b), variability in sensor properties such as impedance or wiring quality, noise sources, and biological factors such as cognitive state or medications. Existing neural decoding algorithms are poorly suited to handle variability in recording condition, resulting in intermittent performance and a need for frequent decoder retraining[4,5,13,17].

The clinical viability of BMIs would be much improved by making decoders robust to recording condition changes[18,19], and several recent studies have begun to focus on this problem. We can broadly divide the conditions that a BMI will encounter into one of two types: 1) conditions that are completely different from what has been previously encountered, and 2) conditions that share some commonality with ones previously encountered. For existing BMI methods, both of these situations necessitate some interruption of function while the decoder is updated to handle the new condition. One strategy for minimizing this interruption is to use adaptive decoders, which update their parameters based on new data collected during the BMI's use (rather than collecting new training data for a *de novo* decoder) to try to better match the new recording condition [4,10,20–29]. In the first case, this is likely the best that can be done. But in the second case, BMI interruption could in principle be avoided altogether by a decoder capable of exploiting the similarities between the current and previously encountered conditions (Fig. 1c).

We were motivated to try this complimentary strategy because chronic BMI systems do typically encounter recording conditions in which there is some commonality with past recording conditions[8,10,13,14,27,28,30–32]. Furthermore, these systems generate and store months, or even years, of neural and kinematic data as part of their routine use. Almost all of this past data is left unused in existing BMI systems: decoders are trained using the most recently available data, typically from a block of calibration trials at the start of that day's experiment, or from a recent previous experiment[33]. Using this historical data would be difficult for most BMI decoders, as they are linear (e.g.,[2,6]). Linear decoders are prone to underfitting heterogeneous training sets, such as those that might be sampled from months of data. To overcome this limitation, an essential aspect of our approach is to use a nonlinear and computationally 'powerful' decoder (i.e. one capable of approximating any complex, nonlinear dynamical systems), which should be capable of learning a diverse set of neural-to-kinematic mappings.

Specifically, we tested whether one could gain traction on the decoder robustness problem by exploiting this idle wealth of stored data using an artificial recurrent neural network (RNN). We did this with a three-pronged approach. The first was the use of the nonlinear RNN. The second was to train the decoder from many months of previously recorded data. Third, to 'harden' the decoder against being too reliant on any given pattern of inputs, we artificially injected additional variability into the data during decoder training.

The fact that conventional state-of-the-art decoding methods, which tend to be linear or at least of limited computational complexity[34], work well for closed-loop BMI control of 2D cursors demonstrates that the model mismatch of assuming linear neural-to-kinematic mappings is well tolerated for a given recording condition. Nevertheless, when neural-to-kinematic mappings change over time, a conventional decoder trained on many days' data is almost certainly not going to fully benefit from this abundance of data. This is because it requires a powerful nonlinear algorithm to learn a set of different context-dependent mappings, even if these individual mappings from neural firing rates to kinematics were entirely linear (which they are not). Methods such as linear Kalman filters can at best only learn an average mapping, 'splitting the difference' to reduce error across the training set. This approach is not well-suited for most of the recording conditions. We therefore developed a new BMI decoder using a nonlinear RNN variant called the multiplicative recurrent neural network (MRNN) developed by Sutskever and colleagues[35] using their Hessian-free technique for training RNNs[36]. Several properties of the MRNN architecture, which was originally used for character-level language modeling, make it attractive for this neural prosthetic application. First, it is recurrent, and can therefore 'remember' state across time (e.g. during the course of a movement), potentially better matching the time-varying, complex relationships between neural firing rates and kinematics[37,38]. Second, its 'multiplicative' architecture increases computational power by allowing the neural inputs to influence the internal dynamics of the RNN by changing the recurrent weights (Fig. 2a). Loosely speaking, this allows the MRNN to learn a 'library' of different neural-to-kinematic mappings that are appropriate to different recording conditions. The MRNN was our specific choice of non-linear method for learning a variety of neural-to-kinematic mappings, but this general approach is likely to work well with many out-of-the-box RNN variants, such as a standard RNN (e.g. [38]) or LSTM[39]. It is also completely complementary to adaptive decoding.

We evaluated decoders using two non-human primates implanted with chronic multielectrode arrays similar to those used in ongoing clinical trials. We first show that training the MRNN with more data from previous recording sessions improves accuracy when decoding new neural data, and that a single MRNN can be trained to accurately decode hand reach velocities across hundreds of days. We next present closed-loop results showing that an MRNN trained with many days' worth of data is much more robust than a state-of-the-art Kalman filter based decoder (the Feedback Intention Trained Kalman filter "FIT-KF"[40]) to two types of recording condition changes likely to be encountered in clinical BMI use: the unexpected loss of signals from highly-informative electrodes, and day-to-day changes. Finally, we show that this robustness does not come at the cost of reduced performance under more ideal (unperturbed) conditions: in the absence of artificial challenges, the MRNN provides excellent closed-loop BMI performance and slightly outperforms the FIT-KF. To our knowledge, this is the first attempt to improve robustness by using a large and heterogeneous training dataset: we used roughly two orders of magnitude more data than in previous closed-loop studies.

## RESULTS

**An MRNN can leverage large amounts of training data to improve decoder performance**

We first tested whether training the MRNN with many days' worth of data can improve offline decoder performance across a range of recording conditions. This strategy was motivated by our observation that the neural correlates of reaching – as recorded with chronic arrays – showed day-to-day similarities (Supplementary Fig. 1). For a typical recording session, the most similar recording came from a chronologically close day, but occasionally the most similar recording condition was found in chronologically distant data. MRNN decoders were able to exploit these similarities: Fig. 2b shows that as more days' data (each consisting of ~500 point to point reaches) were used to train the decoder, the accuracy of reconstructing reach velocities of test datasets increased (positive correlation between number of training days and decoded velocity, $r^2 = 0.24$ for monkey R, $r^2 = 0.20$ for monkey L, $p < 0.001$ for both). Two results from this analysis were particularly encouraging. First, using more training data substantially increased the decode accuracy for the 'hard' days that challenged decoders trained with only a few days' data (e.g., test day 51 for monkey R). Second, this improvement did not come at the cost of worse performance on the initially 'easy' test days. These results demonstrate that larger training datasets better prepare the MRNN for a variety of recording conditions, and that learning to decode additional recording conditions did not diminish the MRNN's capability to reconstruct kinematics under recording conditions that it had already 'mastered'. There was not a performance versus robustness trade-off.

We then tested whether the MRNN's computational capacity could be pushed even further by training it using data from 154 (250) different days' recording sessions from monkey R (L), which spanned 22 (34) months (Fig. 2c). Across all these recording sessions' held-out test trials, the MRNN's offline decode accuracy was $r^2 = 0.81 \pm 0.04$ (mean ± s.d., monkey R) and $r^2 = 0.84 \pm 0.03$ (monkey L). For comparison, we tested the decode accuracy of the FIT-KF trained in two ways: either specifically using reaching data from that particular day ("FIT Sameday"), or trained on the same large multiday training dataset ("FIT Long"). Despite the multitude of recording conditions that the MRNN had to learn, on every test day each monkey's single MRNN outperformed that day's FIT Sameday filter (monkey R: FIT-Sameday $r^2 = 0.57 \pm 0.05$, $p < 0.001$ signed-rank test comparing all days' FIT-Sameday and MRNN $r^2$; monkey L: $r^2 = 0.52 \pm 0.05$, $p < 0.001$). Unsurprisingly, a linear FIT-KF did not benefit from being trained with the same large multiday training set and also performed worse than the MRNN (monkey R: FIT Long $r^2 = 0.56$, $p < 0.001$ comparing all days' FIT Long to MRNN $r^2$; monkey L: $r^2 = 0.46 \pm 0.05$, $p < 0.001$).

We note that, while these offline results demonstrate that the MRNN can learn a variety of recording conditions, experiments are required to evaluate whether this type of training leads to increased decoder robustness under closed-loop BMI cursor control. In closed-loop BMI cursor control, the BMI user updates his or her motor commands as a result of visual feedback, resulting in distributions of neural activity that are different than that of the training set. Thus, results from offline simulation and closed-loop BMI control may differ [32,41–43]. To this end, we next report closed-loop experiments that demonstrate the benefit of this training approach.

**Robustness to unexpected loss of the most informative electrodes**

We next performed closed-loop BMI cursor control experiments to test the MRNN's robustness to recording condition changes. The first set of experiments challenged the decoder with an unexpected loss of inputs from multiple electrodes. The MRNN was trained with a large corpus of hand reaching training data up through the previous day's session (119 - 129 training days for monkey R, 212 - 230 days for monkey L). Then, its closed-loop performance was evaluated on a Radial 8 Task while the selected electrodes' input firing rates were artificially set to zero. By changing how many of the most informative electrodes were dropped ('informative' as determined by their mutual information with reach direction; see Methods), we could systematically vary the severity of the challenge. Since this experiment was meant to simulate sudden failure of electrodes during BMI use (after the decoder had already been trained), we did not retrain or otherwise modify the decoder based on knowledge of which electrodes were dropped. There were no prior instances of these dropped electrode sets having zero firing rates in the repository of previously collected training data (Supplementary Fig. 2). Thus, this scenario is an example of an unfamiliar recording condition (zero firing rates on the dropped electrodes) having commonality with a previously encountered condition (the patterns of activity on the remaining electrodes).

We found that the MRNN was robust to severe electrode dropping challenges. It suffered only a modest loss of performance after losing up to the best 3 (monkey R) or 5 (monkey L) electrodes (Fig. 3). We compared this to the electrode-dropped performance of a FIT-KF decoder trained with hand reaching calibration data from the beginning of that day's experiment[6,40] ("FIT Sameday") by alternating blocks of MRNN and FIT Sameday control in an "ABAB" interleaved experiment design. FIT Sameday decoder's performance worsened dramatically when faced with this challenge. Across all electrodes-dropped conditions, Monkey R acquired 52% more targets per minute using the MRNN, while Monkey L acquired 92% more targets. Supplementary Video 2 shows a side-by-side comparison of the MRNN and FIT-KF decoders with the 3 most informative electrodes dropped.

Although the past datasets used to train the MRNN never had these specific sets of highly important electrodes disabled, our technique of artificially perturbing the true neural activity during MRNN training did generate training examples with reduced firing rates on various electrodes (as well as examples with increased firing rates). The MRNN had therefore been broadly trained to be robust to firing rate reduction on subsets of its inputs. Subsequent closed-loop comparisons of MRNN electrode dropping performance with and without this training data augmentation confirmed its importance (Supplementary Fig. 3a). An additional offline decoding simulation, in which MRNN decoders were trained with varying dataset sizes with and without training data augmentation, further shows that both the MRNN architecture and its training data augmentation are important for robustness to electrode dropping (Supplementary Fig. 4). These analyses also suggest that large training dataset size does not impart robustness to these particular recording condition changes, which is not surprising given that the previous datasets did not include examples of these electrodes being dropped.

**Robustness to naturally occurring recording condition changes**

The second set of closed-loop robustness experiments challenged the MRNN with naturally occurring day-to-day recording condition changes. In contrast to the highly variable recording conditions encountered in human BMI clinical trials, neural recordings in our laboratory setup are stable within a day and typically quite stable on the timescale of days (Supplementary Fig. 1 and [10]). Therefore, in order to challenge the MRNN and FIT-KF decoders with greater recording condition variability, we evaluated them after withholding the most recent several months of recordings from the training data. We refer to this many month interval between the most recent training data day and the first test day as the training data "gap" in these "stale training data" experiments. The gaps were chosen arbitrarily within the available data, but to reduce the chance of outlier results, we repeated the experiment with two different gaps for each monkey.

For each gap, we trained the MRNN with a large dataset consisting of many months of recordings preceding the gap and compared it to two different types of FIT-KF decoders. The "FIT Old" decoder was trained from the most recent available training day (i.e. the day immediately preceding the gap); this approach was motivated under the assumption that the most recent data were most likely to be similar to the current day's recording condition. The "FIT Long" decoder was trained from the same multiday dataset used to train the MRNN and served as a comparison in which a conventional decoder is provided with the same quantity of data as the MRNN. The logic underlying this FIT Long approach was that despite the Kalman filter being ill-suited for fitting multiple heterogeneous datasets, this 'averaged' decoder might still perform better than the FIT-KF Old trained using a single distant day.

We found that the MRNN was the only decoder that was reliably usable when trained with stale data (Fig. 4). FIT Old performed very poorly in both monkeys, failing completely (defined as the monkey being unable to complete a block using the decoder, see Methods) in 4/6 monkey R experimental sessions and 6/6 monkey L sessions. FIT Long performed better than FIT Old, but its performance was highly variable – it was usable on some test days but failed on others. In Monkey R, the across-days average acquisition rate was 105% higher for the MRNN than FIT Long ($p < 0.01$, paired *t*-test). Monkey L's MRNN did not perform as consistently well as Monkey R's, but nevertheless demonstrated a trend of outperforming FIT Long (32% improvement, $p = 0.45$), in addition to decidedly outperforming FIT Old, which failed every session. Although monkey L's FIT Long outperformed the MRNN on one test day, on all other test days FIT Long was either similar to, or substantially worse than, MRNN. Moreover, whereas the MRNN could be used to control the cursor every day, FIT Long was not even capable of acquiring targets on some days. Further tests of additional FIT Old decoders confirmed that they generally perform poorly (Supplementary Fig. 5). The lack of consistent usability by any of the FIT-KF decoders (Old or Long) demonstrates that having access to a large repository of stale training data does not enable training a single Kalman filter that is robust to day-to-day variability in recording conditions. In contrast, an MRNN trained with this large dataset was consistently usable.

To further demonstrate the consistency of these results, we performed offline simulations in which we tested MRNN decoders on additional sets of training and test datasets separated by a gap. Each set was non-overlapping with the others, and together they spanned a wide range of each animal's research career. We observed the same trends in these offline simulations: MRNNs trained with many previous days of training data outperformed FIT Old and FIT Long decoders (Supplementary Fig. 6). In these analyses we also dissected which components of our decoding strategy contributed to the MRNN's robustness. We did this by comparing MRNNs trained with varying numbers of days preceding the gap, with or without training data spike rate perturbations. The results show that training using more data, and to a lesser extent incorporating data augmentation (see also closed-loop comparisons in Supplementary Fig. 3b), contributed to the MRNN's robustness to naturally occurring recording condition changes.

**High performance BMI using the MRNN decoder**
Finally, we note that the MRNN's robustness to challenging recording conditions did not come at the cost of reduced performance under more 'ideal' conditions, i.e. without electrode dropping or stale training data. During the electrode dropping experiments, we also evaluated the MRNN's closed-loop performance after being trained using several months' data up through the previous day. In this scenario, the MRNN enabled both monkeys to accurately and quickly control the cursor. Supplementary Movie 1 shows example cursor control using the MRNN. This data also allowed us to compare the MRNN's performance to that of a FIT Sameday decoder in back-to-back "ABAB" tests. Figure 5a shows representative cursor trajectories using each decoder, as well as under hand control. Figure 5b shows that across nine experimental sessions and 4,000+ trials with each decoder, Monkey R acquired targets 7.3% faster with the MRNN (619 ± 324 ms mean ± s.d. vs. 668 ± 469 ms, p < 0.01, rank-sum test). Monkey L acquired targets 10.8% faster with the MRNN (743 ± 390 ms vs. 833 ± 532 ms, p < 0.01, rank-sum test) across 8 sessions and 2,500+ trials using each decoder. These online results corroborate the offline results presented in Fig. 2c; both show that an MRNN trained from many days' recording conditions outperforms the FIT Kalman filter trained from training data collected at the start of the experimental session.

A potential risk inherent to a computationally powerful decoder such as the MRNN is that it will overtrain to the task structure of the training data and fail to generalize to other tasks. Most of our MRNN training data were from arm reaches on a Radial 8 Task similar to the task used for evaluation (albeit with 50% further target distance). We therefore also tested whether the MRNN enabled good cursor control on the Random Target Task, in which the target could appear in any location in a 20 x 20 cm workspace (Supplementary Fig. 7). Monkey R performed the Random Target Task on two experimental sessions and averaged a 99.4% success rate, with mean distance-normalized time to target of 0.068 s/cm. Monkey L performed one session of this task at a 100% success rate with mean normalized time to target of 0.075 s/cm. To provide context for these metrics, we also measured Random Target Task performance using arm control. Monkey R's arm control success rate was 100%, with 0.055 s/cm mean normalized time to target, during the same experimental sessions as his MRNN Random Target Task data. Monkey L's arm control success rate was 97.7%, with 0.055 s/cm mean normalized time to target, during one session several days following his MRNN test.

**DISCUSSION**

We developed the MRNN decoder to help address a major problem hindering the clinical translation of BMIs: once trained, decoders can be quickly rendered ineffective due to recording condition changes. A number of complementary lines of research are aimed at making BMIs more robust, including improving sensors to record from more neurons more reliably (e.g.,[44]); decoding multiunit spikes[10,30,45] or local field potentials[31,32,46] which appear to be more stable control signals than single unit activity; and using adaptive decoders that update their parameters to follow changing neural-to-kinematic mappings [4,10,20–29,47]. Here we present the MRNN as a proof-of-principle of a new and different approach: build a fixed decoder whose architecture allows it to be trained to be inherently robust to recording condition changes it has previously encountered as well as to new conditions that have some similarity to a previously encountered condition.

We stress that all of these approaches are complementary in several respects. For example, a decoder that is inherently more robust to neural signal changes, such as the MRNN, would still benefit from improved sensors, could operate on a mix of input signal types including single- and multiunit spikes and field potentials, and is especially well positioned to benefit from decoder adaptation. When performance degrades due to recording condition changes, both supervised[10,21–23,25,27,29] and unsupervised[4,20,24,26] adaptive decoders need a period of time in which control is at least good enough that the algorithm can eventually infer the user's intentions and use these to update its neural-to-kinematic model. Improved robustness may "buy enough time" to allow the decoder's adaptive component to rescue performance without interrupting prosthesis use. Here we've demonstrated the MRNN's advantages over a state-of-the-art static decoder, but comparing this strategy both against and together with adaptive decoding remains a future direction.

We demonstrated the MRNN's robustness to two types of recording condition changes. These changes were chosen because they capture key aspects of the changes that commonly challenge BMI decoders during clinical use. The stale training data experiments showed that the MRNN was usable under conditions where the passage of time would typically require recalibration of conventional decoders such as the FIT-KF. We do not mean to suggest that in a clinical setting one would want to – or would often have to – use a BMI without any training data from the past several months. Rather, we used this experimental design to model recording condition changes that can happen on the timescale of hours in human BMI clinical trials[13]. Possible reasons for the greater recording condition variability observed in human participants compared to non-human primates include: more movement of the array relative to the human brain due to larger cardiovascular pulsations and epidural space; greater variability in the state of the BMI user (health, medications, fatigue, cognitive state); and more electromagnetic interference from the environment. The MRNN can take advantage of having seen the effects of these sources of variability in previously accumulated data; it can therefore be expected to become more robust over time as it builds up a "library" of neural-to-kinematic mappings under different recording conditions.

The electrode dropping experiments, which demonstrated the MRNN's robustness to an unexpected loss of high-importance electrodes, are important for two reasons. Firstly, sudden loss of input signals (e.g., due to a electrode connection failure[48,49]), is a common BMI failure mode that can be particularly deleterious to conventional BMI decoders[50]. The MRNN demonstrates considerable progress in addressing this so-called "errant unit" problem. Secondly, these results demonstrate that the MRNN trained with artificially perturbed neural data can be relatively robust even to a recording condition change that has not been encountered in past recordings.

The MRNN's robustness did not come at the cost of diminished performance under more ideal conditions. This result is nontrivial given the robustness-focused decisions that went into its design (e.g. perturbing the input spike trains in the training set). Instead, we found that the MRNN was excellent under favorable conditions, slightly outperforming a state-of-the-art same day trained FIT-KF decoder. Taken together, these results demonstrate that the MRNN exhibits robustness to a variety of clinically relevant recording condition changes, without sacrificing peak performance. These advances may help to reduce the onerous need for clinical BMI users to collect frequent retraining data.

One disadvantage of this class of nonlinear decoders trained from large datasets, when compared to traditional linear decoders trained on smaller datasets, is the longer training time required to fit their more numerous parameters. In the present study, which we did not optimize for fast training, this took multiple hours. This could be substantially sped up by iteratively updating the decoder with new data instead of retraining *de novo* and by leveraging faster computation available with graphics processing units, parallel computing, or custom hardware. A second disadvantage of the MRNN is that it appears to require more training data to saturate its performance (Fig. 2b) compared to conventional methods, such as FIT-KF, that are trained from sameday calibration data. We do not view this as a major limitation because the motivation for using the MRNN is to take advantage of accumulated previous recordings. Nonetheless, it will be valuable to compare the present approach with other decoder architectures and training strategies, which may yield similar performance and robustness while requiring less training data.

The MRNN decoder's robustness was due to the combination of a large training data corpus, deliberate perturbation of the training data, and a computationally powerful architecture that was able to effectively learn this diverse training data. While it may seem obvious that successfully learning more training data is better, this is not necessarily true. Older data only helps a decoder if some of these past recordings capture neural-to-kinematic relationships that are similar to that of the current recording condition. Our offline and closed-loop MRNN robustness results suggest that this was indeed the case for the two monkeys used in this study. While there are indications that this will also be true in human BMI studies[14], validating this remains an important future question. The relevance of old data to present recording conditions also enables a different robustness-enhancing approach: store a library of different past decoders and evaluate each to find a decoder well-suited for the current conditions (e.g.,[10]). However, since offline analyses are poor predictors of closed-loop performance[32,42,45,51], this approach necessitates a potentially lengthy decoder selection process. Using a single decoder (such as the MRNN) that works across many recording conditions avoids switching-related downtime.

In addition to training with months of previous data, we improved the MRNN's robustness by intentionally perturbing the training neural data. In the present study we applied random Gaussian firing rate scaling based on a general assumption that the decoder should be broadly robust to both global and private shifts in observed firing rates. This perturbation type proved effective, but we believe that this approach (called data augmentation in the machine learning community) can potentially be much more powerful when combined with specific modeling of recording condition changes that the experimenter wants to train robustness against. For example, data augmentation could incorporate synthetic examples of losing a particularly error-prone set of electrodes; recording changes predicted by models of array micro-movement or degradation; and perhaps even the predicted interaction between kinematics and changes in cognitive state or task context. We believe this is an important avenue for future research.

We view the success of our specific MRNN decoder implementation as a validation of the more general BMI decoder strategy of training a computationally powerful nonlinear decoder to a large quantity of data representing many different recording conditions. This past data need not have been collected explicitly for the purpose of training as was done in this study; neural data and corresponding kinematics from past closed-loop BMI use can also serve as training data[4,10]. It is likely that other nonlinear decoding algorithms will also benefit from this strategy, and that there are further opportunities to advance the reliability and performance of BMIs by starting to take advantage of these devices' ability to generate large quantities of data as part of their regular use.

## METHODS

**Animal model and neural recordings**

All procedures and experiments were approved by the Stanford University Institutional Animal Care and Use Committee. Experiments were conducted with adult male rhesus macaques (R and L, ages 8 and 18, respectively), implanted with 96-electrode Utah arrays (Blackrock Microsystems Inc., Salt Lake City, UT) using standard neurosurgical techniques. Monkeys R and L were implanted 30 months and 74 months prior to the primary experiments, respectively. Monkey R had two electrode arrays implanted, one in caudal dorsal premotor cortex (PMd) and the other in primary motor cortex (M1), as estimated visually from anatomical landmarks. Monkey L had one array implanted on the border of PMd and M1. Within the context of the simple point-to-point arm and BMI reach behavior of this study, we observed qualitatively similar response properties between these motor cortical areas; this is consistent with previous reports of a gradient of increasing preparatory activity, rather than stark qualitative differences, as one moves more rostral from M1 [52–56]. Therefore, and in keeping with standard BMI decoding practices [6,8,10,24,38,40,46], we did not distinguish between M1 and PMd electrodes.

Behavioral control and neural decode were run on separate PCs using the xPC Target platform (Mathworks, Natick, MA), enabling millisecond-timing precision for all computations. Neural data were initially processed by Cerebus recording system(s) (Blackrock Microsystems Inc., Salt Lake City, UT) and were available to the behavioral control system within 5 ± 1 ms. Spike counts were collected by applying a single negative threshold, set to -4.5 times the root mean square of the spike band of each electrode. We decoded "threshold crossings", which contain spikes from one or more neurons in the electrode's vicinity, as per standard practice for intracortical BMIs [1,4,6,7,10,15,16,31,38,40] because threshold crossings provide roughly comparable population-level velocity decode performance to sorted single-unit activity, without time-consuming sorting[30,45,57–59], and may be more stable over time[30,45]. To orient the reader to the quality of the neural signals available during this study, Supplementary Data 1 provides statistics of several measures of electrodes' 'tuning' and cross-talk.

**Behavioral tasks**

We trained the monkeys to acquire targets with a virtual cursor controlled by either the position of the hand contralateral to the arrays or directly from neural activity. Reaches to virtual targets were made in a 2D fronto-parallel plane presented within a 3D environment (MSMS, MDDF, USC, Los Angeles, CA) generated using a Wheatstone stereograph fused from two LCD monitors with refresh rates at 120 Hz, yielding frame updates within $7 \pm 4$ ms [43]. Hand position was measured with an infrared reflective bead tracking system at 60 Hz (Polaris, Northern Digital, Ontario, Canada). During BMI control, we allowed the monkey's reaching arm to be unrestrained[47,60] so as to not impose a constraint upon the monkey that during BMI control he must generate neural activity that does not produce overt movement[61].

In the *Radial 8 Task* the monkey was required to acquire targets alternating between a center target and one of eight peripheral targets equidistantly spaced on the circumference of a circle. For our closed-loop BMI experiments, the peripheral targets were positioned 8 cm from the center target. In hand reaching datasets used for decoder training and offline decode, the targets were either 8 cm or 12 cm (the majority of datasets) from the center. In much of Monkey L's training data, the three targets forming the upper quadrant were placed slightly further (13 and 14 cm) based on previous experience that this led to decoders with improved ability to acquire targets in that quadrant. To acquire a target, the monkey had to hold the cursor within a 4 cm × 4 cm acceptance window centered on the target for 500 ms. If the target was acquired successfully, the monkey received a liquid reward. If the target was not acquired within 5 s (BMI control) or 2 s (hand control) of target presentation, the trial was a failure and no reward was given.

Although the data included in this study's decoder training datasets and offline analyses span many months of each animal's research career, these data start after each animal was well-trained in performing point-to-point planar reaches; day-to-day variability when making the same reaching movements was modest. To quantify behavioral similarity across the study, we took advantage of having collected the same "Baseline Block" task data at the start of most experimental sessions: 171/185 monkey R days, 398/452 monkey L days. This consisted of ~200 trials of arm-controlled Radial 8 Task reaches, with targets 8 cm from the center. For each of these recording sessions, we calculated the mean hand x- and y-velocities (averaged over trials to/from a given radial target) throughout a 700 ms epoch following radial target onset for outward reaches and 600 ms following center target onset for inward reaches (inward reaches were slightly faster). We concatenated these velocity time series across the 8 different targets, producing 10,400 ms x-velocity and y-velocity vectors from each recording session. Behavioral similarity between any two recording sessions was then measured by the Pearson correlation between the datasets' respective x- and y- velocity vectors. Then, the two dimensions' correlations were averaged to produce a single correlation value between each pair of sessions. These hand velocity correlations were 0.90 ± 0.04 (mean ± s.d. across days) for monkey R, and 0.91 ± 04 for monkey L.

We measured closed-loop BMI performance on the Radial 8 Task using two metrics. Target acquisition rate is the number of peripheral targets acquired divided by the duration of the task. This metric holistically reflects cursor control ability because, unlike time to target, it is negatively affected by failed trials and directly relates to the animal's rate of liquid reward. Targets per minute is calculated over all trials of an experimental condition (i.e., which decoder was used) and therefore yields a single measurement datum per day/experimental condition. Across-days distributions of a given decoder's targets per minute performance were consistent with a normal distribution (Kolmogorov-Smirnov test), justifying our use of paired *t*-tests statistics when comparing this metric. This is consistent with the measure reflecting the accumulated outcome of many hundreds of random processes (individual trials). As a second measure of performance that is more sensitive when success rates are high and similar between decoders (such as the "ideal" conditions where we presented no challenges to the decoders), we compared times to target. This measure consists of the time between when the target appeared and when the cursor entered the target acceptance window prior to successfully acquiring the target, but does not include the 500 ms hold time (which is constant across all trials). Times to target are only measured for successful trials to peripheral targets, and were only compared when success rates were not significantly different (otherwise, a poor decoder with a low success rate that occasionally acquired a target quickly by chance could nonsensically "outperform" a good decoder with 100% success rate but slower times to target). Because these distributions were not normal, we used the Mann–Whitney–Wilcoxon rank-sum tests when comparing two decoders' times to target.

In the *Random Target Task* each trial's target appeared at a random location within a 20 cm × 20 cm region centered within a larger workspace that was 40 x 30 cm. A new random target appeared after each trial regardless of whether this trial was a success or a failure due to exceeding the 5 s time limit. The target location randomization enforced a rule that the new target's acceptance area could not overlap with that of the previous target. Performance on the Random Target Task was measured by success rate (the number of successfully acquired targets divided by the total number of presented targets) and the normalized time to target. Normalized time to target is calculated for successful trials following another successful trial, and is the duration between target presentation and target acquisition (not including the 500 ms hold time), divided by the straight-line distance between this target's center and the previously acquired target's center [62].

**Decoder comparison experiment design**

All offline decoding comparisons between MRNN and FIT-KF were performed using test data that was held out from the data used to train the decoders. Thus, although the MRNN has many more parameters than FIT-KF, both of these fundamentally different algorithm types were trained according to best practices with matched training data. This allows their performance to be fairly compared.

When comparing online decoder performance using BMI-controlled Radial 8 Target or Random Target Tasks, the decoders were tested in using an interleaved "block-set" design in which contiguous ~200 trial blocks of each decoder were run followed by blocks of the next decoder, until the block-set comprising all tested decoders was complete and the next block-set began. For example, in the electrode dropping experiments (Figure 3), this meant an "AB AB" design where A could be a block of MRNN trials and B could be a block of FIT Sameday trials. For the stale training data experiments (Figure 4), an "ABCD ABCD ABCD… " design was used to test the four different decoders. When switching decoders, we gave the monkey ~20 trials to transition to the new decoder before starting 'counting' performance in the block; we found this to be more than sufficient for both animals to adjust. For electrode dropping experiments, the order of decoders within each block-set was randomized across days. For stale training data experiments, where several decoders often performed very poorly, we manually adjusted the order of decoders within block-sets so as to keep the monkeys motivated by alternating what appeared to be more and less frustrating decoders. All completed blocks were included in the analysis. Throughout the study, the experimenters knew which decoder was in use, but all comparisons were quantitative and performed by the same automated computer program using all trials from completed blocks. The monkeys were not given an overt cue to the decoder being used.

During online experiments, we observed that when a decoder performed extremely poorly, such that the monkey could not reliably acquire targets within the 5 second time limit, the animal stopped performing the task before the end of the decoder evaluation block. To avoid frustrating the monkeys, we stopped a block if the success rate fell below 50% after at least 10 trials. This criterion was chosen based on pilot studies in which we found that below this success rate, the monkey would soon thereafter stop performing the task and would frequently refuse to re-engage for a prolonged period of time. Our interleaved block design meant that each decoder was tested multiple times on a given experimental session, which in principle provides the monkey multiple attempts to finish a block with each decoder. In practice, we found that monkeys could either complete every block or no blocks with a given decoder, and we refer to decoders that could not be used to complete a block as having "failed". These decoders' performance was recorded as 0 targets per minute for that experimental session. The exception to the above was that during an electrode dropping experiment session, we declared both FIT-KF Sameday and MRNN as having failed for a certain number of electrodes dropped if the monkey could not complete a block with either decoder. That is, we did not continue with a second test of both (unusable) decoders as per the interleaved block design, because this would have unduly frustrated the animal.

We performed this study with two monkeys, which is the conventional standard for systems neuroscience and BMI experiments using a non-human primate model. No monkeys were excluded from the study. We determined how many experimental sessions to perform as follows. For all offline analyses, we examined the dates of previous experimental sessions with suitable arm reaching data and selected sets of sessions with spacing most appropriate for each analysis (e.g., closely spaced sessions for Figure 2b, all of the available data for Figure 2c, two clusters with a gap for stale training analyses). All these pre-determined sessions were then included in the analysis. For the stale training data experiments (Figure 4), the choice of two gaps with three test days each was pre-established. For the electrode dropping experiments (Figure 3), we did not know *a priori* how electrode dropping would affect performance and when each decoder would fail. We therefore determined the maximum number of electrodes to drop during the experiment and adjusted the number of sessions testing each drop condition during the course of experiments to comprehensively explore the "dynamic range" across which decoder robustness appeared to differ. For both of these experiments, during an experimental session additional block-sets were run until the animal became satiated and disengaged from the task. We did not use formal effect size calculations to make data sample size decisions, but did perform a variety of experiments with large numbers of decoder comparison trials (many tens of thousands) so as to be able to detect substantial decoder performance differences. For secondary online experiments (Supplementary Figures 3 and 7), which served to support offline analyses (Supplementary Figure 3) or demonstrate that the MRNN could acquire other target locations (Supplementary Figure 7), we chose to perform only 1-3 sessions per animal in the interest of conserving experimental time.

**Neural decoding using a Multiplicative Recurrent Neural Network (MRNN)**
At a high level, the MRNN decoder transforms inputs $\mathbf{u}(t)$, the observed spike counts on each electrode at a particular time, into a cursor position and velocity output. This is accomplished by having first trained the weights of an artificial recurrent neural network such that when the network is provided a time series of training neural data inputs, the training data kinematic outputs can be accurately 'read out' from this neural network's state. The rest of this section will describe the architecture, training, and use of the MRNN for the purpose of driving a BMI.

The generic recurrent network model is defined by an $N$-dimensional vector of activation variables, $\mathbf{x}$, and a vector of corresponding "firing rates", $\mathbf{r} = \tanh \mathbf{x}$. Both $\mathbf{x}$ and $\mathbf{r}$ are continuous in time and take continuous values. In the standard RNN model, the input affects the dynamics as an additive time-dependent bias in each dimension. In the MRNN model, the input instead directly parameterizes the artificial neural network's recurrent weight matrix, allowing for a multiplicative interaction between the input and the hidden state. One view of this multiplicative interaction is that the hidden state of the recurrent network is selecting an appropriate decoder for the statistics of the current dataset. The equation governing the dynamics of the activation vector is of the form suggested in [35], but adapted in this study to continuous time in order to control the smoothness to MRNN outputs,

$$\tau \dot{\mathbf{x}}(t) = -\mathbf{x}(t) + \mathbf{J}^{\mathbf{u}(t)} \mathbf{r}(t) + \mathbf{b}^x.$$

The $N \times N \times |\mathbf{u}|$ tensor $\mathbf{J}^{\mathbf{u}(t)}$ describes the weights of the recurrent connections of the network, which are dependent on the $E$-dimensional input, $\mathbf{u}(t)$. The symbol $|\mathbf{u}|$ denotes the number of unique values $\mathbf{u}(t)$ can take. Such a tensor is unusable for continuous valued $\mathbf{u}(t)$ or even discrete valued $\mathbf{u}(t)$ with prohibitively many values. To make these computations tractable, the input is linearly combined into $F$ factors and $\mathbf{J}^{\mathbf{u}(t)}$ is factorized [35] according to the following formula:

$$\mathbf{J}^{\mathbf{u}(t)} = \mathbf{J}^{xf} \cdot \mathrm{diag}\left(\mathbf{J}^{fu}\mathbf{u}(t)\right) \cdot \mathbf{J}^{fx},$$

where $\mathbf{J}^{xf}$ has dimension $N{\times}F$, $\mathbf{J}^{fu}$ has dimension $F{\times}E$, $\mathbf{J}^{fx}$ has dimension $F{\times}N$, and $\mathrm{diag}(\mathbf{v})$ takes a vector, $\mathbf{v}$, and returns a diagonal matrix with $\mathbf{v}$ along the diagonal. One can directly control the complexity of interactions by choosing $F$. Additionally, the network units receive a bias $\mathbf{b}^x$. The constant $\tau$ sets the time scale of the network, so we set $\tau$ on the order of hundreds of milliseconds to allow meaningful interactions. The output of the network is read out from a weighted sum of the network firing rates plus a bias, defined by the equation

$$\mathbf{z}(t) = \mathbf{W_o}\,\mathbf{r}(t) + \mathbf{b}^z,$$

where $\mathbf{W_o}$ is an $M{\times}N$ matrix, and $\mathbf{b}^z$ is an $M$-dimensional bias.

|  | Monkey R | Monkey L |
|---|---|---|
| $\Delta t$ | 20 ms | 20-30 ms |
| $\tau$ | 100 ms | 100-150 ms |
| $N$ | 100 | 50 |
| $F$ | 100 | 50 |
| $\sigma_{trial}$ | 0.045 | 0.045 |
| $\sigma_{electrode}$ | 0.3 | 0.3 |
| $g_{xf}$ | 1.0 | 1.0 |
| $g_{fu}$ | 1.0 | 1.0 |
| $g_{fx}$ | 1.0 | 1.0 |
| $E$ | 192 | 96 |
| days of training data | 82-129 | 189-230 |
| years spanned | 1.59 | 2.77 |
| # params in each MRNN | 39502 | 9952 |
| $\beta$ | 0.99 | 0.99 |

**Table 1. Network and training parameters used for the closed-loop MRNN BMI decoder**

## MRNN training

We began decoder training by instantiating MRNNs of network size $N = 100$ (monkey R) and $N = 50$ (monkey L) with $F = N$ in both cases (see Table 1 for all key parameters). For monkey R, who was implanted with two multielectrode arrays, $E = $ 192, while for monkey L with one array, $E = 96$. The non-zero elements of the non-sparse matrices $\mathbf{J}^{xf}, \mathbf{J}^{fu}, \mathbf{J}^{fx}$ are drawn independently from a Gaussian distribution with zero mean and variance $g_{xf}/F, g_{fu}/E$, and $g_{fx}/N$, with $g_{xf}, g_{fu}$, and $g_{fx}$ set to 1.0 in this study. The elements of $\mathbf{W_0}$ are initialized to zero, and the bias vectors $\mathbf{b}^x$ and $\mathbf{b}^z$ are also initialized to 0.

The input $\mathbf{u}(t)$ to the MRNN (through the matrix $\mathbf{J}^{\mathbf{u}(t)}$) is the vector of binned spikes at each time step. Concatenating across time in a trial yields training data matrix, $\mathbf{U}^j$, of binned spikes of size $E \times T^j$, where $T^j$ is the number of times steps for the $j^{\text{th}}$ trial. Data from five consecutive actual monkey-reaching trials are then concatenated together to make one "MRNN training" trial. The first two actual trials in an MRNN training trial were used for seeding the hidden state of the MRNN (i.e., not used for learning), whereas the next three actual trials were used for learning. With the exception of the first two actual trials from a given recording day, the entire set of actual trials are used for MRNN learning by incrementing the actual trial index that begins each training trial by one.

The parameters of the network were trained offline to reduce the averaged squared error between the measured kinematic training data and the output of the network, $\mathbf{z}(t)$. Specifically, we used the Hessian-Free (HF) optimization method[36,63] for RNNs (but adapted to the continuous-time MRNN architecture). HF is an exact 2$^{\text{nd}}$ order method that uses back-propagation through time to compute the gradient of the error with respect to the network parameters. The set of trained parameters is $\{\mathbf{J}^{xf}, \mathbf{J}^{fu}, \mathbf{J}^{fx}, \mathbf{b}^x, \mathbf{W_0}, \mathbf{b}^z\}$. The HF algorithm has three critical parameters: the minibatch size, the initial lambda setting, and the max number of conjugate gradient iterations. We set these parameters to one-fifth the total number of trials, 0.1, and 50, respectively. The optimizations were run for 200 steps and a snapshot of the network was saved every 10 steps. Among these snapshots, the network with the lowest cross-validation error on held-out data was used in the experiment.

We independently trained two separate MRNN networks to each output a 2-dimensional ($M = 2$) signal, $\mathbf{z}(t)$. The first network learned to output the normalized hand position through time in both the horizontal ($x$) and vertical ($y$) spatial dimensions. The second MRNN learned to output the hand velocity through time, also in the $x$ and $y$ dimensions. We calculated hand velocities from the positions numerically using central differences.

In this study we trained a new MRNN whenever adding new training data; this allowed us to verify that the training optimization consistently converged to a high-quality decoder. However, it is easy to iteratively update an MRNN decoder with new data without training from scratch. By adding the new data to the training corpus and using the existing decoder weights as the training optimization's initial conditions, the MRNN will more rapidly converge to a new high-quality decoder.

**Improving the MRNN by training with many datasets and intentionally perturbed inputs**

A critical element of achieving both high performance and robustness in the MRNN decoder was training the decoder using data from many previous recording days spanning many months. When training datasets included data from more than one day, we randomly selected a small number of trials from each day for a given minibatch. In this way, every minibatch of training data sampled the input distributions from all training days.

A second key element of training robustness to recording condition changes was a form of data augmentation in which we intentionally introduced perturbations to the neural spike trains that were used to train the MRNN. The concatenated input, $\widehat{\mathbf{U}} = [\mathbf{U}^i, ..., \mathbf{U}^{i+4}]$ was perturbed by adding and removing spikes from each electrode. We focus on electrode $c$ of the $j^{\text{th}}$ training trial, i.e., a row vector of data $\widehat{\mathbf{U}}_{c,:}^j$. Let the number of actual observed spikes in $\widehat{\mathbf{U}}_{c,:}^j$ be $n_c^j$. This number was perturbed according to

$$\hat{n}_c^j = \eta^j \eta_c n_c^j,$$

where both $\eta^j$ and $\eta_c$ are Gaussian variables with a mean of one and standard deviations of $\sigma_{\text{trial}}$ and $\sigma_{\text{electrode}}$, respectively. Conceptually, $\eta^j$ models a global firing rate modulation across all electrodes of the array (e.g., array movement, arousal), while $\eta_c$ models electrode by electrode perturbations such as electrode dropping or moving baselines in individual neurons. If $\hat{n}_c^j$ was less than zero or greater than $2n_c^j$, it was resampled, which kept the average number of perturbed spikes in a given electrode and training trial roughly equal to the average number of true (unperturbed) spikes in the same electrode and training trial. Otherwise, if $\hat{n}_c^j$ was greater than $n_c^j$, then $\hat{n}_c^j - n_c^j$ spikes were added to random time bins of the training trial. If $\hat{n}_c^j$ was less than $n_c^j$, then $n_c^j - \hat{n}_c^j$ spikes were randomly removed from time bins of the training trial that already had spikes. Finally, if $\hat{n}_c^j = n_c^j$, nothing was changed.

The process of perturbing the binned spiking data occurred anew on every iteration of the optimization algorithm, i.e. in the HF algorithm, the perturbation $\hat{n}_c^j = \eta^j \eta_c n_c^j$ occurs after each update of the network parameters. Note that these input data perturbations were only applied during MRNN training; when the MRNN was used for closed-loop BMI control, true neural spike counts were provided as inputs. Supplementary Figure 3 shows the closed-loop control quality difference between the MRNN trained with and without this data augmentation.

**Controlling a BMI cursor with MRNN network output**

Once trained, the MRNNs were compiled into the embedded real-time operating system and run in closed-loop to provide online BMI cursor control. The decoded velocity and position were initialized to 0, as was the MRNN hidden state. Thereafter, at each decode time step the parallel pair of MRNNs received binned spike counts as input and had their position and velocity outputs blended to yield a position estimate. This was used to update the drawn cursor position. The on-screen position that the cursor moves to during BMI control, $d_x(t), d_y(t)$, is defined by

$$d_x(t) = \beta(d_x(t - \Delta t) + \gamma_v v_x(t - \Delta t)\Delta t) + (1 - \beta)\gamma_p p_x(t)$$
$$d_y(t) = \beta(d_y(t - \Delta t) + \gamma_v v_y(t - \Delta t)\Delta t) + (1 - \beta)\gamma_p p_y(t)$$

where $v_x, v_y, p_x, p_y$ are the normalized velocity and positions in the *x* and *y* dimensions and $\gamma_v, \gamma_p$ are factors that convert from the normalized velocity and position, respectively, to the coordinates of the virtual-reality workspace. The parameter $\beta$ sets the amount of position versus velocity decoding and was set to 0.99. In effect, the decode was almost entirely dominated by velocity, with a slight position contribution to stabilize the cursor in the workplace (i.e., offset accumulated drift). Note that when calculating offline decode accuracy (Fig. 2), we set $\beta$ to 1 to more fairly compare the MRNN to the FIT-KF decoder, which decodes velocity only.

We note that although 1) the MRNN's recurrent connections mean that previous inputs affect how subsequent near-term inputs are processed, and 2) our standard procedure was to retrain the MRNN with additional data after each experimental session, the MRNN is *not* an "adaptive" decoder in the traditional meaning of the term. Its parameters are fixed during closed-loop use, and therefore when encountering recording condition changes, the MRNN cannot "learn" from this new data to update its neural-to-kinematic mappings in the way that adaptive decoders do (e.g., [4,24,27]). Insofar as its architecture and training regime make the MRNN robust to input changes, this robustness is "inherent" rather than "adaptive."

**Neural decoding using a Feedback Intention Trained Kalman Filter (FIT-KF)**

We compared the performance of the MRNN to FIT-KF [40]. The FIT-KF is a Kalman filter where the underlying kinematic state, $\mathbf{z}(t)$, comprises the position and velocity of the cursor as well as a bias term. Observations of the neural binned spike counts, $\mathbf{y}(t)$, are used to update the kinematic state estimate. With $\Delta t$ denoting bin width (25 ms in this study), the FIT-KF assumes the kinematic state gives rise to the neural observations according to the following linear dynamical system:

$$\mathbf{z}(t + \Delta t) = \mathbf{A}\mathbf{z}(t) + \mathbf{w}(t)$$
$$\mathbf{y}(t) = \mathbf{C}\mathbf{z}(t) + \mathbf{q}(t)$$

where $\mathbf{w}(t)$ and $\mathbf{q}(t)$ are zero-mean Gaussian noise with covariance matrices $\mathbf{W}$ and $\mathbf{Q}$ respectively. The Kalman filter is a recursive algorithm that estimates the state $\mathbf{z}(t)$ using the current observation $\mathbf{y}(t)$ and the previous state estimate $\mathbf{z}(t - \Delta t)$. Previous studies have used such decoders to drive neural cursors (e.g. [5,38,64]).

The parameters of this linear dynamical system, $\mathbf{A}, \mathbf{W}, \mathbf{C}, \mathbf{Q}$, are learned in a supervised fashion from hand reach training data as previously reported[6,65]. The FIT-KF then incorporates two additional innovations. First, it performs a rotation of the training kinematics using the assumption that at every moment in time, the monkey intends to move the cursor directly towards the target. Second, it assumes that at every time step, the monkey has perfect knowledge of the decoded position via visual feedback. This affects Kalman filter inference in two ways: first, the covariance of the position estimate in Kalman filtering is set to 0, and secondly, the neural activity that is explainable by the cursor position is subtracted from the observed binned spike counts. These innovations are further described in [6,40].

**Mutual information for determining electrode dropping order**

When testing the decoders' robustness to unexpected electrode loss, we determined which electrodes to drop by calculating the mutual information between each electrode's binned spike counts and the reach direction. This metric produced a ranking of electrodes in terms of how statistically informative they were of the reach direction; importantly, this metric is independent of the decoder being used. Let $p$ denote the distribution of an electrode's binned firing rates, $y$ denote the binned spike counts lying in a finite set $Y$ of possible binned spike counts, $M$ denote the number of reach directions, and $x_j$ denote reach direction $j$. The set $Y$ comprised {0,1,2,3,4,5+} spike counts, where any spike counts greater than or equal to 5 were counted towards the same bin ("5+", corresponding to an instantaneous firing rate of 250 Hz in a 20 ms bin). We calculated the entropy of each electrode,

$$H(Y) = -\sum_{y \in Y} p(y) \log p(y),$$

as well as its entropy conditioned on the reach direction

$$H(Y|X) = -\sum_{j=1}^{M} p(x_j) \sum_{y \in Y} p(y|x_j) \log p(y|x_j).$$

From these quantities, we calculated the mutual information between the neural activity and the reach direction as $I_{drop}(X;Y) = H(Y) - H(Y|X)$. We dropped electrodes in order from highest to lowest mutual information.

**Principal angles of neural subspaces analysis**

For a parsimonious scalar metric of how similar patterns of neural activity during reaching were between a given pair of recording days (used in Supplementary Fig. 1), we calculated the minimum principal angle between the neural subspaces of each recording day. We defined the neural subspace on a recording day as the top $K$ principal components of the neural coactivations. Put more simply, we asked how similar day $i$ and day $j$'s motifs of covariance between electrodes' activity were during arm reaching. Specifically, we started with a matrix $Y_i$ from each day $i$ consisting of neural activity collected while the monkey performed ~200 trials of a Radial 8 Task (8 cm distance to targets) using arm control; this task has been run at the start of almost every experimental session conducted using both monkeys R and L since array implantation. $Y_i$ is of dimensionality $E \times T$, where E is the number of electrodes and T is the number of non-overlapping 20 ms bins comprising the duration of this task. We next subtracted from each row of $Y_i$ that electrode's across-days mean firing rate (we also repeated this analysis without across-days mean subtraction and observed qualitatively similar results, not shown). To obtain the principal components, we performed eigenvalue decomposition on the covariance matrix $Y_i Y_i^T$ (note, $Y_i$ is zero mean), and defined the matrix $V_i$ as the first $K$ eigenvectors. $V_i$ had dimensions $E \times K$, where each column $k$ is the vector of principal component coefficients (eigenvector) corresponding to the $k^{th}$ largest eigenvalue of the decomposition. Supplementary Fig. 1 was generated using $K = 10$, i.e., keeping the first 10 PCs, but the qualitative appearance of the data was similar when $K$ was varied from 2 to 30 (not shown). Finally, the difference metric between days $i$ and $j$ was computed as the minimum of the $K$ subspace angles between matrices $\boldsymbol{V}_i$ and $\boldsymbol{V}_j$. Subspace angles were computed using the *subspacea* MATLAB function[66].

**Data and code availability**

All relevant data and analysis code can be made available from the authors upon request.

**Acknowledgments**

We thank Mackenzie Risch, Michelle Wechsler, Liana Yates, Shannon Smith, and Rosie Steinbach for surgical assistance and veterinary care; Evelyn Castaneda and Beverly Davis for administrative support; Boris Oskotsky for information technology support. This work was supported by the National Science Foundation Graduate Research Fellowship (J.C.K., S.D.S.); NSF IGERT 0734683 (S.D.S.); Christopher and Dana Reeve Paralysis Foundation (S.I.R. and K.V.S.); and the following to K.V.S.: Burroughs Welcome Fund Career Awards in the Biomedical Sciences, Defense Advanced Research Projects Agency Reorganization and Plasticity to Accelerate Injury Recovery N66001-10-C-2010, US National Institutes of Health Institute of Neurological Disorders and Stroke Transformative Research Award R01NS076460, US National Institutes of Health Director's Pioneer Award 8DP1HD075623-04, US National Institutes of Health Director's Transformative Research Award (TR01) from the NIMH #5R01MH09964703,

**Author contributions**

D.S., S.D.S., and J.C.K. designed the experiments. D.S. trained the MRNN decoders. S.D.S. and J.C.K. conducted the experiments and data analysis. S.D.S. wrote the manuscript. J.C.K., D.S., and S.I.R. assisted in manuscript preparation. S.I.R. was responsible for surgical implantation. K.V.S. was involved in all aspects of experimentation, data review, and manuscript preparation.

**Competing financial interests**

The authors declare no competing financial interests.

# FIGURES (with legends below)

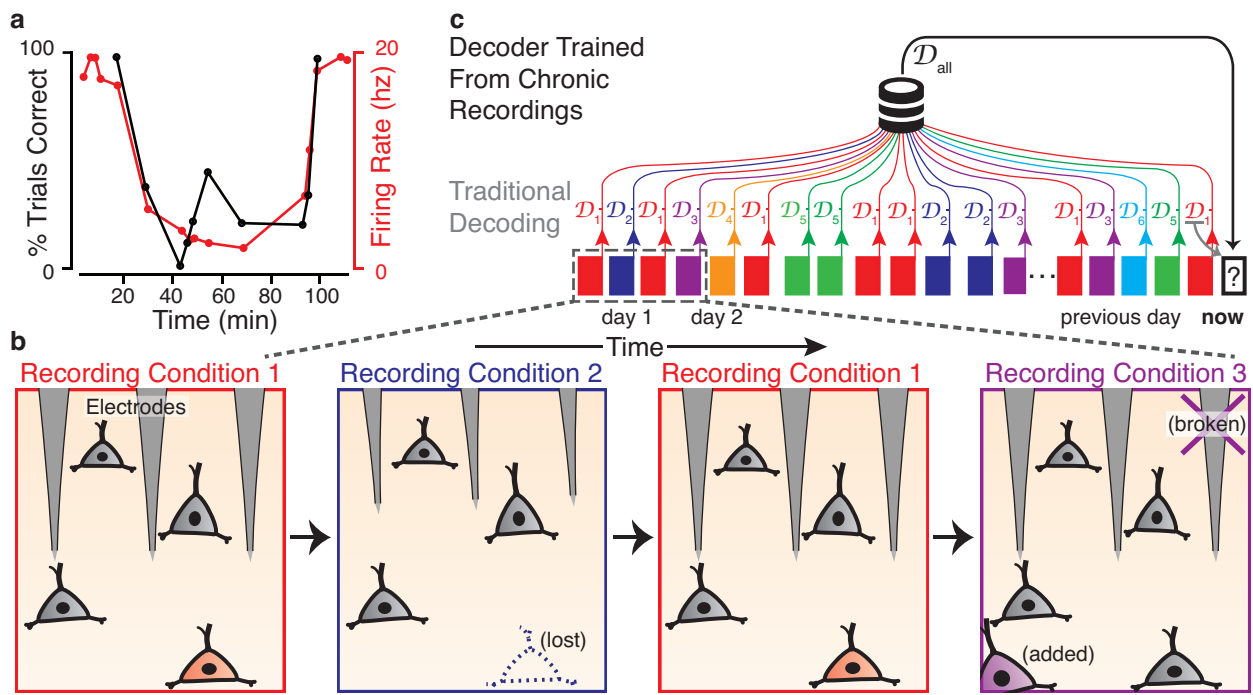

**Figure 1. Our strategy for training a decoder robust to recording condition changes**

(**a**) Example data from a BMI clinical trial showing sudden decoder failure caused by a recording condition change. The black trace shows the participant's closed-loop performance over the course of an experiment using a fixed Kalman filter. An abrupt drop in performance coincides with a reduction in the observed firing rate (red trace) of a neuron with a high decoder weight. Both the neuron's firing rate and decoder performance spontaneously recover ~40 minutes later. Adapted from figure 7 of [13]. (**b**) A cartoon depicting one hypothetical cause of the aforementioned change: micro-motion of the electrodes leads to Recording Condition 2, in which spikes from the red-shaded neuron are lost. BMI recovery corresponds to a shift back to Condition 1. Over time, further changes will result in additional recording conditions; for example, Condition 3 is shown caused by a disconnected electrode and an additional neuron entering recording range. (**c**) Recording conditions (schematized by the colored rectangles) will vary over the course of chronic intracortical BMI use. We hypothesize that oftentimes new conditions are similar to ones previously encountered (repeated colors). Typically, decoders are fit from short blocks of training data and are only effective under that recording condition (decoders $\mathcal{D}_1$, $\mathcal{D}_2$, …). Consider the goal of training a decoder for use at time "now" (black rectangle on right). Standard practice is to use decoder $\mathcal{D}_1$ trained from the most recently available data (e.g. from the previous day or the start of the current experiment). $\mathcal{D}_1$ will perform poorly if the recording condition encountered differs from its training data. To increase the likelihood of having a decoder that will perform well given the current recording condition, we tested a new class of decoder, $\mathcal{D}_{\text{all}}$, trained using a large collection of previous recording conditions.
Page 43

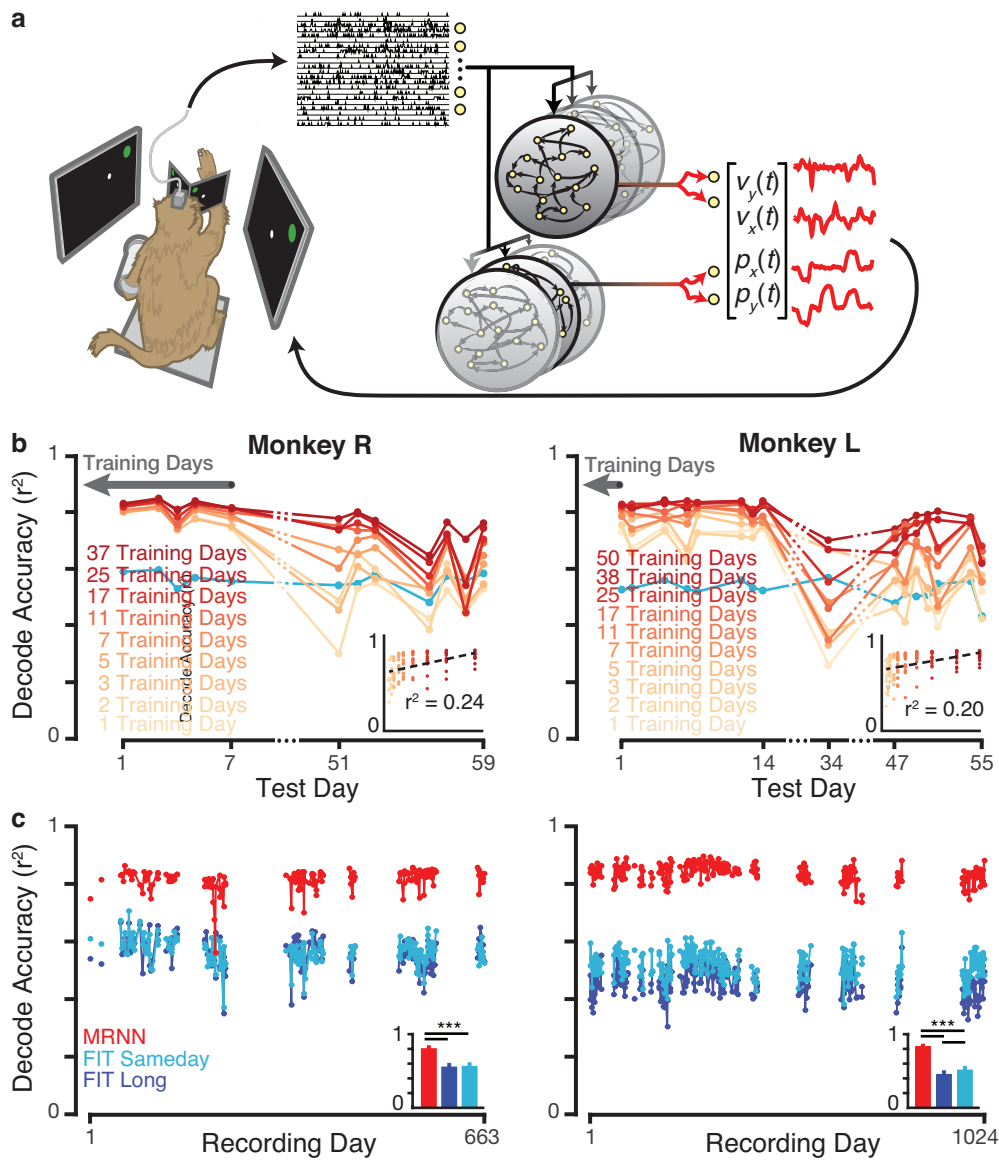

**Figure 2. An MRNN decoder can harness large training datasets**

(**a**) Schematic of the experimental setup and MRNN decoder. A monkey performed a target acquisition task using his hand while multiunit spikes were recorded from multielectrode arrays in motor cortex. Data from many days of this task were used to train two MRNNs such that velocity and position could be read out from the state of their respective internal dynamics. These internal dynamics are a function of the binned spike counts; thus, the MRNN can conceptually be thought of as selecting which internal dynamics rules are active at any given time based on the input signals. During closed-loop use, the decoded velocity and position outputs were blended together to continuously update the on-screen cursor. (**b**) The MRNN was better prepared for future days' recording conditions after being trained with more previously collected data. We evaluated its ability to reconstruct offline hand velocity on 12 (16) monkey R (L) test days after training with increasing numbers of previous days' datasets. Training data

Page 44

were added by looking further back in time so as to not conflate training data recency with data corpus size. In monkey R the early test days also contributed training data (with the test trials held out). In monkey L, from whom more suitable data was available, the training datasets start with the day prior to the first test day. Thus, except for the few monkey R overlap days, all training data comes chronologically before the test days. More training data (darker colored traces) improved decode accuracy, especially when decoding more chronologically distant recording conditions (later test days). For comparison, we also plotted performance of a traditional decoder (FIT Kalman filter) trained from each individual day's training data ("FIT Sameday", light blue). (Insets) show the same MRNN data in a scatter plot of decode accuracy versus number of training days (99 data points for monkey R, 160 for L), with linear fit trend lines revealing a significant positive correlation. (**c**) An MRNN could successfully learn even larger datasets spanning many more recording days. An MRNN (red trace) was trained with data from 154 (250) monkey R (L) recording days spanning many months. Its offline decoding accuracy on held-out trials from each of these same days was compared to that of the FIT Sameday (light blue). We also tested a single FIT-KF trained using the same large dataset as the MRNN ("FIT Long", dark blue). Gaps in the connecting lines denote recording gaps of more than ten days. (Insets) mean ± s.d. decode accuracy across all recording days. Stars denote $p < 0.001$ differences (signed-rank test). The MRNN outperformed both types of FIT-KF decoders on every day's dataset.

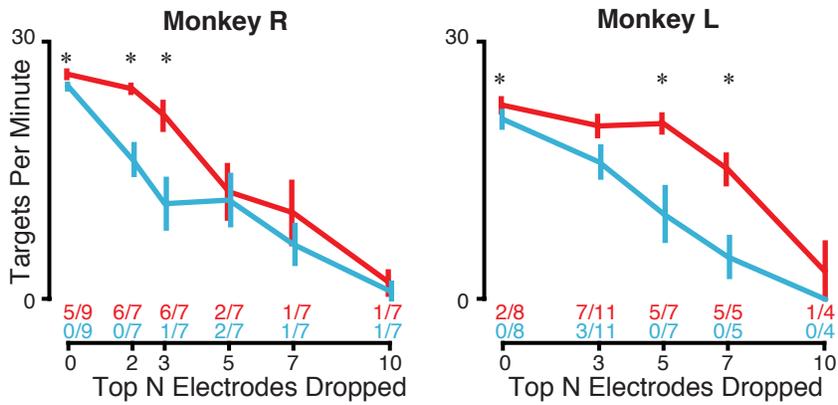

**Figure 3. Robustness to unexpected loss of the most important electrodes**

Closed-loop BMI performance using the MRNN (red) and FIT Sameday (blue) decoders while simulating an unexpected loss of up to 10 electrodes by setting the firing rates of these electrodes to zero. The mean and s.e.m. across experimental sessions' targets per minute performance is shown for each decoder as a function of how many electrodes were removed. Stars denote conditions for which the MRNN significantly outperformed FIT Sameday across sessions ($p < 0.05$, paired *t*-test). The text above each condition's horizontal axis tick specifies for how many of the individual evaluation days MRNN (red fraction) or FIT Sameday (blue fraction) performed significantly better according to single-session metrics of success rate and time to target. Electrode-dropping order was determined by the mutual information between that electrode's spike count and target direction during arm-controlled reaches.

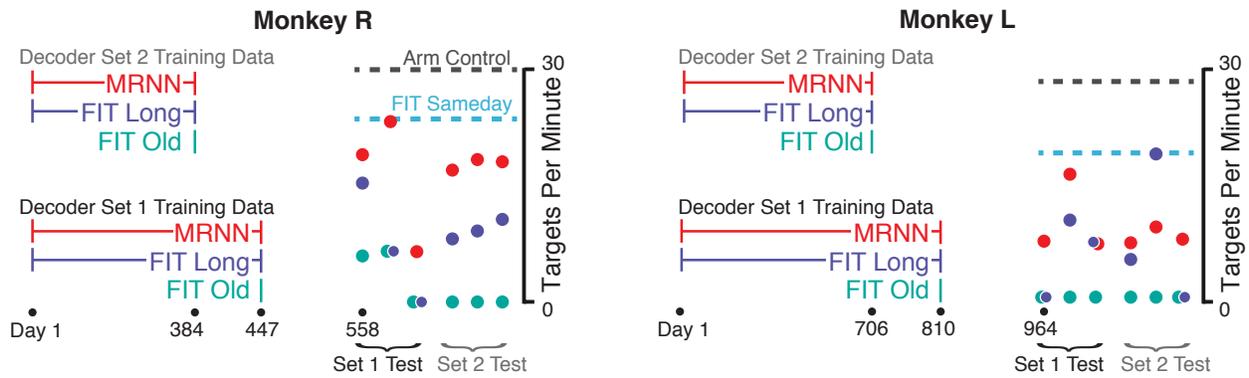

**Figure 4. Robustness to naturally occurring recording condition changes**

Robustness to a natural sampling of neural input variability. We created decoder evaluation conditions in which the neural inputs were likely to be different from much of the training data by withholding access to the most recent several months of data. Each circle corresponds to the mean closed-loop BMI performance using these "stale" MRNN (red), FIT Long (dark blue), and FIT Old (teal) decoders when evaluated on six different experiment days spanning 7 (13) days in monkey R (L). Each test day, these three decoders, as well as a FIT Sameday decoder trained from that day's arm reaches, were evaluated in an interleaved block design. The legend bars also denote the time periods from which training data for each stale decoder came from. We repeated the experiments for a second set of decoders to reduce the chance that the results were particular to the specific training data gap chosen. The training data periods contained 82 and 92 datasets (monkey R), and 189 and 200 training datasets (monkey L). The only decoder that was consistently usable, i.e. did not fail on any test days, was the MRNN. To aid the interpretation of these stale decoder performances, we show the average performance across the six experiment days using arm control (gray dashed line) or a FIT Sameday decoder (blue dashed line).

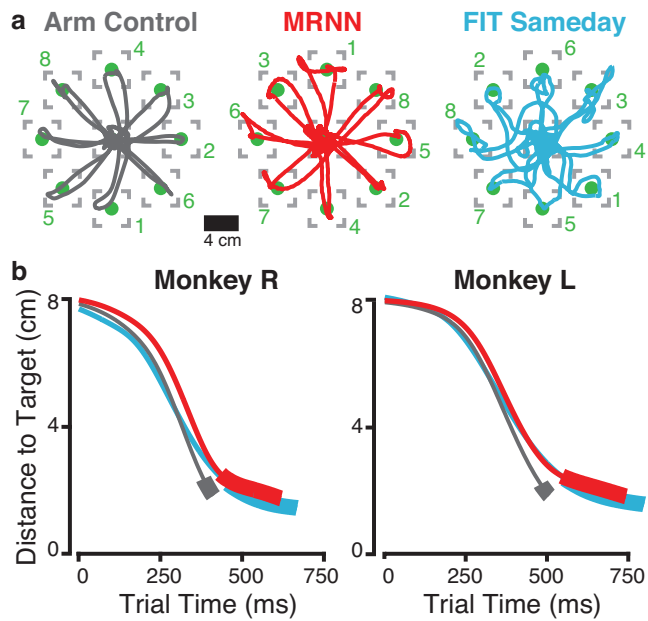

**Figure 5. MRNN has high performance under "ideal" conditions**

(**a**) We compared cursor control using the MRNN (red) trained from many datasets up through the previous day to the FIT Sameday (blue) trained from data collected earlier the same day, without any artificial challenges (i.e. no electrodes dropped or stale training data). Cursor trajectories are shown for eight representative and consecutive center-out-and-back trials of the Radial 8 Task. Gray boxes show the target acquisition area boundaries, and the order of target presentation is denoted with green numbers. For comparison, cursor trajectories under arm control are shown in gray. From dataset R.2014.04.03.

(**b**) Mean distance to target, across all Radial 8 Task trials under these favorable conditions, as a function of trial time using each cursor control mode. Thickened portions of each trace correspond to "dial-in time", i.e. the mean time between the first target acquisition and the final target acquisition. These MRNN and FIT Sameday data correspond to the drop 0 electrodes condition data in Figure 3, and include 4,094 (3,278) MRNN trials and 4119 (3,305) FIT Sameday trials over 9 (8) experimental days in Monkey R (L).

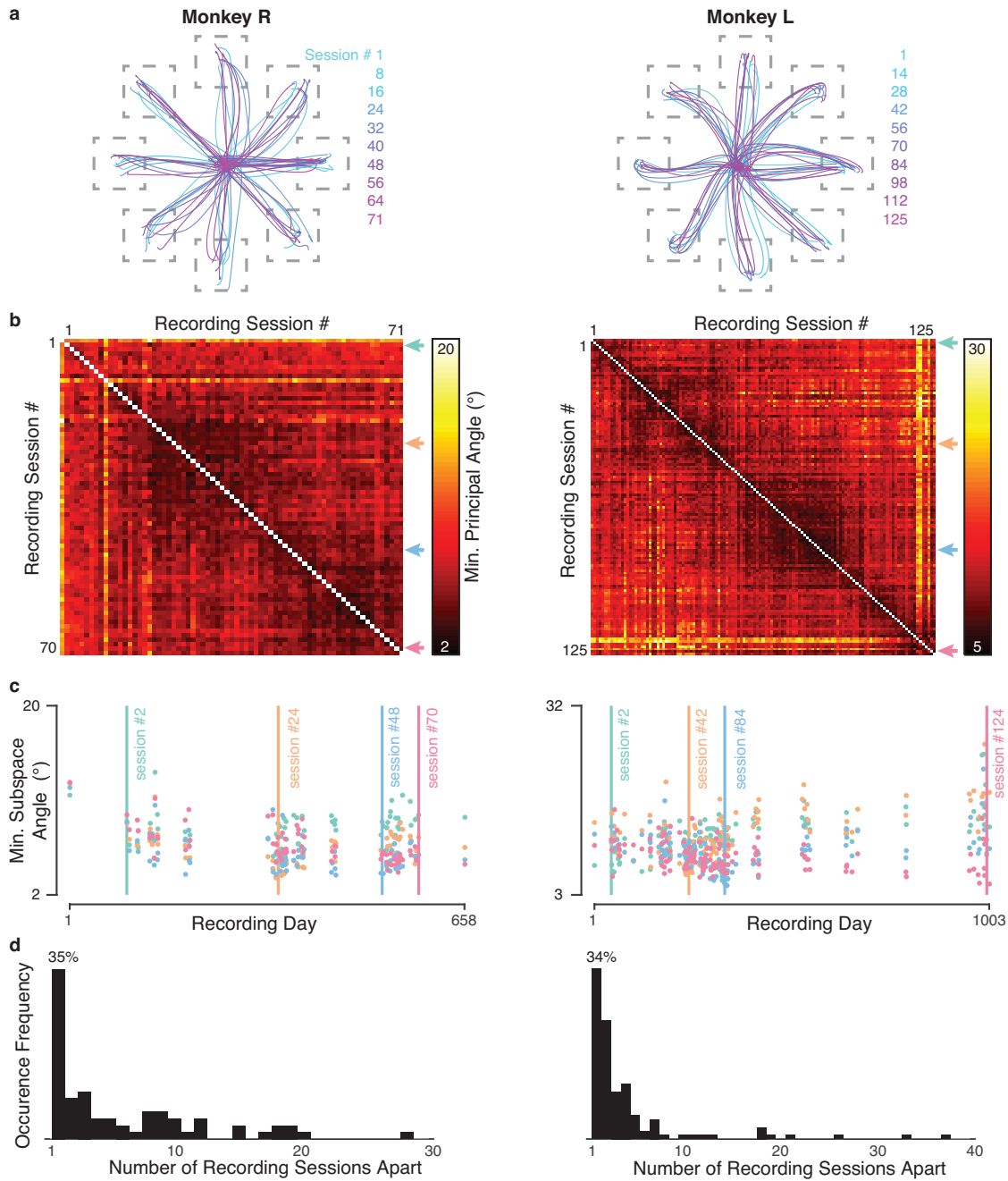

**Supplementary Figure 1. For a given day, similar neural recording conditions can be found on some other day(s).** Chronologically close days tend to have more similar neural recordings, but for a given day there are occasional similar recordings from more distant days. (**a**) To minimize the potential effect of behavioral variability on neural variability, we restricted this analysis to recording sessions with very consistent Radial 8 Target task behavior. Hand velocity correlations between all pairs of sessions within the included set were at least 0.9. Representative hand position traces (mean over trials towards each target) are shown for ten sessions spanning the months analyzed. (**b**) Between-day variability of the structure of neural activity recorded during reaches over the course of many months (71 recording sessions over a 658 day period in monkey R, and 125 sessions spanning 1003 days in monkey L; these correspond to a subset of the days included in Fig. 2c). The color at the intersection of row *i*

and column *j* corresponds to how differently the observed neural activity covaried during recording sessions *i* and *j*. Specifically, we have plotted the minimum principal angle between subspaces spanned by the top 10 eigenvectors of each day's mean-activity-subtracted covariance matrix (see Methods). These 10 eigenvectors captured on average 51 (46)% of single-trial variance for monkeys R (L). Sharp "block" structure transitions typically correspond to a long (many weeks') interval between consecutive recording sessions. (**c**) Four slices through each monkey's principal angle matrix show that for these example days, there were similar recording conditions on other days both near and far in time. Each color series shows the minimum principal angle between every recording day's data and the reference day marked with this color's arrow in panel b. Note that the horizontal axis, which spans the same time range as in panel b, is now in units of (calendar) days rather than session number. Each series' reference day is marked with a vertical line. (**d**) Histograms showing the distribution, across each monkey's recordings, of how many recording sessions apart (either forward or back in time) we observed the most similar neural correlates of reaching as measured by minimum principal angle.

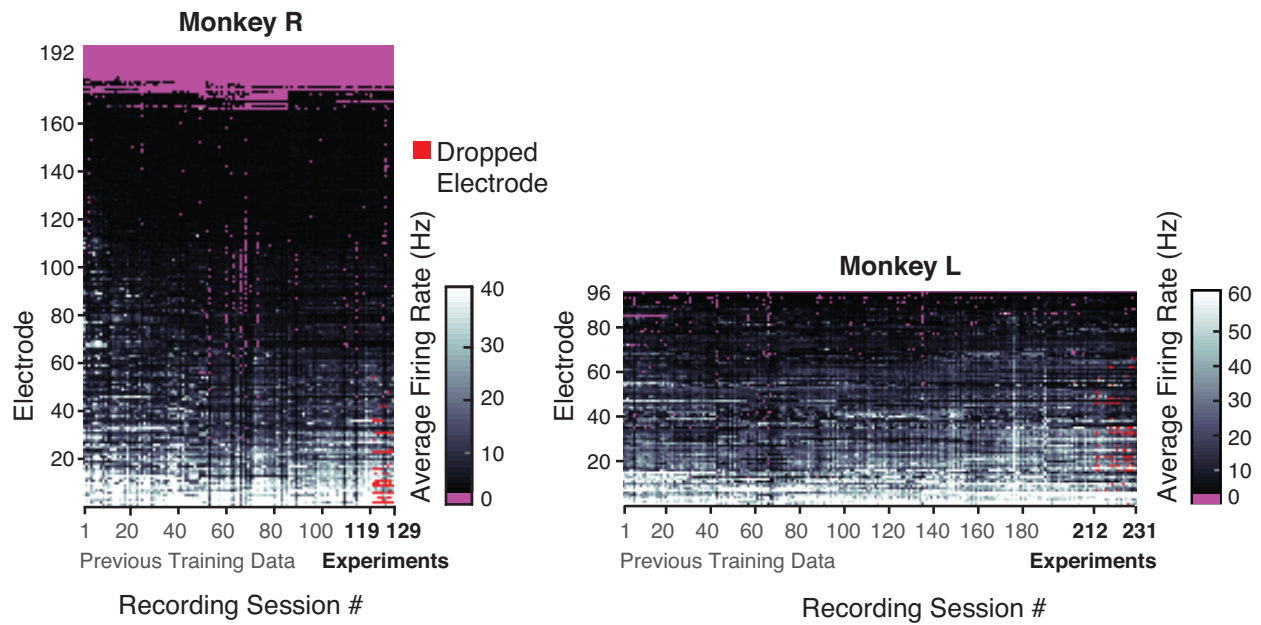

**Supplementary Figure 2. Artificially dropped electrodes were active in the training data.** These plots show each electrode's average firing rate during each dataset used to train the MRNN; electrodes are ordered by descending average firing rate across all recording sessions. Recording sessions numbered in gray were only used for training data. The electrode dropping experiments (Fig. 3) were conducted during the sessions numbered in black. Zero firing rates (i.e. non-functional electrodes) are shown in purple for emphasis, while electrodes selected for dropping on a particular day are shown in red (note that although on a given test session we evaluated different numbers of electrodes dropped, this plot shows each day's broadest dropped set). These dropped electrodes rarely recorded zero firing rates in the training data sessions, and the specific sets of dropped electrodes used to challenge the decoders never all had zero firing rates in the training data.

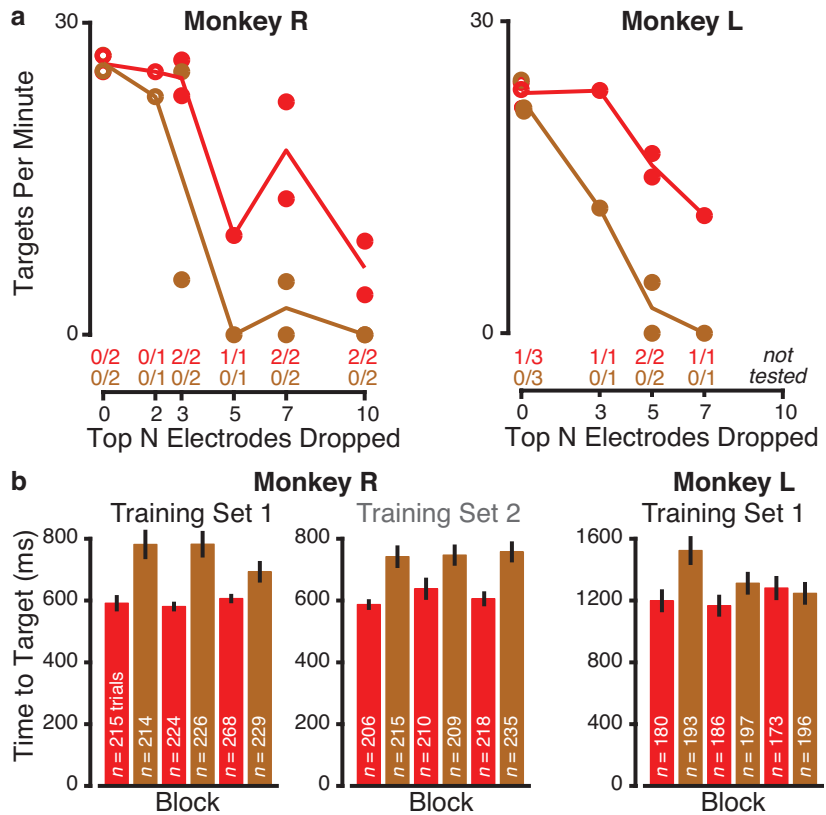

**Supplementary Figure 3. Training data spike rate perturbations improve closed-loop MRNN robustness.**
(**a**) Robustness to electrode dropping. We evaluated the closed-loop BMI performance of the MRNN decoder trained with (red) and without (brown) the spike rate perturbations training data augmentation. Both decoders were evaluated on the same day with firing rates on varying numbers of the most informative electrodes set to zero (similar to Fig. 3). Each circle corresponds to a decoder's targets per minute performance on a given evaluation day. In total there were 3 sessions per monkey. Filled circles denote conditions where there was a significant within-session performance difference between the two decoders tested according to: $p < 0.05$ binomial test on success rate, followed, if success rate was not significantly different, by a more sensitive comparison of times to target ($p < 0.05$, rank-sum test). Fractions above the horizontal axis specify for how many of the individual evaluation days each decoder performed significantly better. Trend lines show the across-sessions mean targets per minute performance for each decoder. The MRNN trained with perturbed firing rates outperformed the MRNN trained without data augmentation when encountering electrode-dropped neural input.
(**b**) Robustness to naturally occurring neural recording condition changes. MRNNs were trained without access to recent training data, as in the Fig. 4 stale training data experiments, either with (red) or without (brown) training data spike rate perturbations. We trained decoders from both of monkey R's stale training data periods and from monkey L's longer stale training data period. Closed-loop BMI performance using these decoders was then compared on the same evaluation day in alternating blocks. Bars show mean ± s.e.m. time to target for each block of trials (success rates using both training paradigms were close to 100%). The MRNN with spike rate perturbations had significantly faster times to target in monkey R ($p < 0.05$, rank-sum test aggregating trials across blocks) and showed a trend of faster times to target in monkey L ($p = 0.066$). Datasets R.2014.03.21 & L.2014.04.04.
Page 51

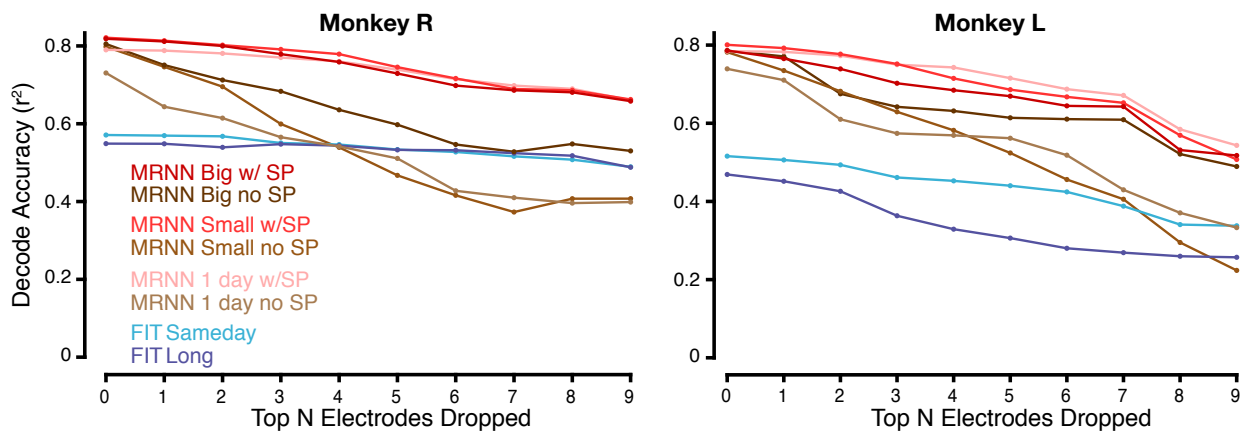

**Supplementary Figure 4. Offline decode experiments testing various training paradigms' robustness to unexpected loss of electrodes**

We performed offline decoding analyses to test how each of the three main components of our method – the use of an MRNN architecture instead of a linear Kalman filter, the use of large training datasets, and the spike count perturbation data augmentation – contributed to improved robustness to unexpected loss of the most important electrodes, similar to the online tests shown in Figure 3. The results suggest that both the use of the MRNN architecture and training data augmentation contributed to the complete system's improved robustness to a novel recording condition consisting of electrode dropping.

      We trained MRNNs with different quantities of data: 1 day of (held out) data from the test day, a 'Small' dataset consisting 10-13 days up to and including the test day, or a "Big" dataset of 40 – 100 days up to and including the test day, with ("w/SP") or without ("no SP") additional spike count perturbations during training.   We also trained a FIT-KF Sameday decoder and a FIT-KF Long which used the same datasets as the MRNN Big datasets. We compared the offline decoding accuracy of each decoder as a function of the number of electrodes dropped, using the same electrode dropping order determination method as in the Figure 3 online experiments. Three decoders were trained for each training paradigm using data from different periods of each monkey's research career; these decoders' training dates correspond to exactly the same as those in Supplementary Figure 6, and each decoder was tested on held out data from its last day of training. We averaged offline hand velocity reconstruction accuracy across each monkey's three testing days.  We found that applying the spike count perturbation always increased MRNN robustness to electrode dropping (compiling $r^2$ across all SP vs all no SP decoders across all three test days, SP decoders performed better than no SP decoders, $p < 0.001$, signed-rank test). Note that, as expected, training with larger dataset sizes did not strongly affect performance or robustness to electrode dropping, since all MRNN's have 'seen' data collected on the same day as the withheld testing data.

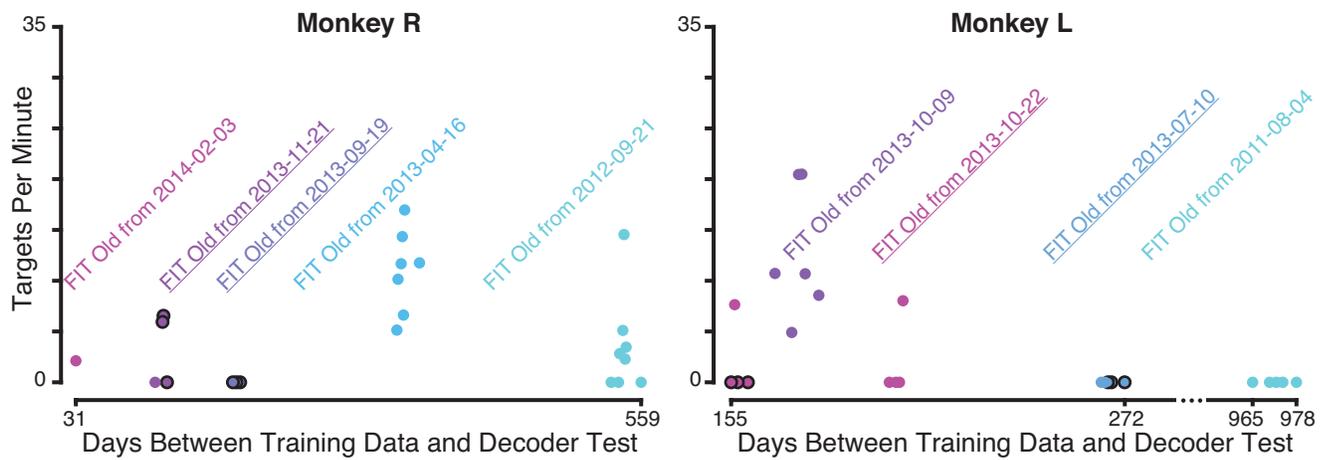

**Supplementary Figure 5. Additional tests showing that FIT Old typically performs poorly.**
We investigated whether our result that three of the four different FIT Old decoders tested in the main stale training data experiments (Fig. 4) failed was due to a particularly unlucky choice of FIT Olds. To better sample the closed-loop performance of FIT-KF decoders trained using old training data, we trained FIT Old decoders from 3 (monkey R) and 2 (Monkey L) additional arbitrarily chosen arm reaching datasets from the monkey's prior experiments. We evaluated all 5 (4) FIT Old decoders on a number of additional days over the course of the current study (8 total test days for monkey R, 13 total test days for monkey L). Each point shows the performance of a particular FIT Old decoder on one test day. Different days' evaluations of the same FIT Old decoder are shown in the same color. Black borders denote data points and black underlines denote decoders that are shared with Fig. 4.

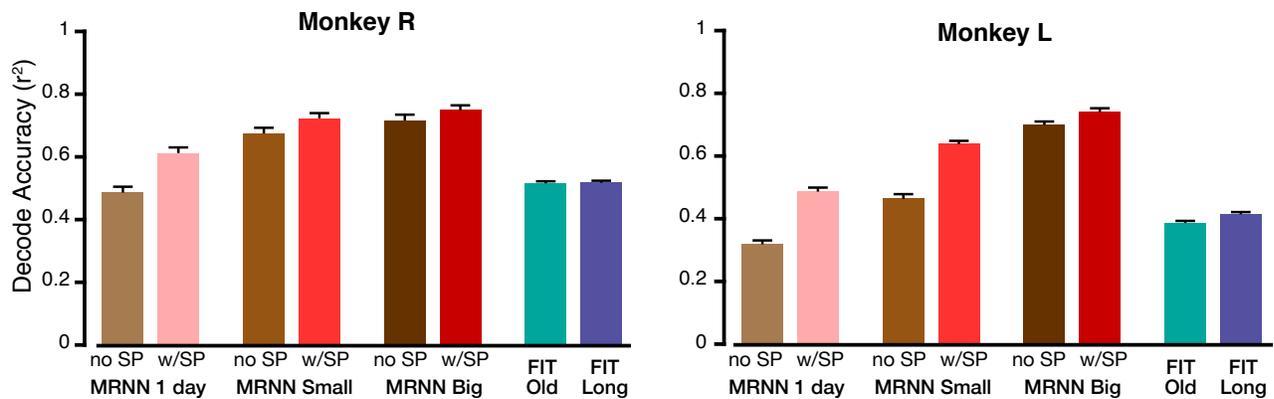

**Supplementary Figure 6. Offline decode experiments testing various training paradigms' robustness to naturally occurring recording condition changes**

We performed offline decoding analyses to test how each of the three main components of our method – the use of an MRNN architecture instead of a linear Kalman filter, the use of large training datasets, and the spike count perturbation data augmentation – contributed to improved robustness to naturally occurring recording condition changes similar to the online 'stale' training data tests shown in Figure 4. We performed offline decoding evaluations across three different training data gaps and found that using more previously collected data and incorporating data augmentation both improved the MRNN's performance on the future test data. MRNNs trained with many datasets outperformed FIT-KFs trained using the same data ("FIT Long") or just the most recent data ("Fit Old"). These results suggest that all three components of the method contributed to the complete system's improved robustness.

(Left) Offline decode results for Monkey R. We performed an offline decode for 8 different types of decoders and aggregated each decoder's performance over the three gaps. The MRNN decoders were either trained with 1 day of data "1 day," a "Small" dataset (gap 1: 13 datasets, gaps 2 and 3: 10 datasets) or a "Big" dataset (gap 1: 40 datasets, gap 2: 44 datasets, gap 3: 37 datasets). The MRNN decoders were also either trained with no spike rate perturbations ("no SP") or with spike rate perturbations ("w/SP"). We also trained a FIT Old using the most recent dataset and a FIT Long which used the same datasets as the MRNN Big datasets. Gap 1 comprised training data from 2012-10-22 (YYYY-MM-DD) to 2013-04-19 and testing data from 2013-07-29 to 2013-11-21 (44 testing days). Gap 2 comprised training data from 2013-07-29 to 2013-11-21 and testing data from 2014-02-03 to 2014-04-07 (37 testing days). Gap 3 comprised training data from 2014-02-03 to 2014-04-07 and testing data from 2014-06-16 to 2014-08-19 (33 testing days). The bars show the mean ± s.d. performance of each training approach across all 3 gaps. We observed the same trends across individual gaps, with the MRNN Big w/SP decoder always achieving the best performance (p < 0.01, signed-rank test with every other decoder, all gaps).

(Right) Same for Monkey L. The Small datasets comprised 10 datasets (gap 1 and 2) or 11 datasets (gap 3), while the Big datasets comprised 103 datasets (gap 1), 105 datasets (gap 2), and 77 datasets (gap 3). Gap 1 comprised training data from 2010-03-04 to 2010-10-26 and testing data from 2011-01-17 to 2011-04-28 (51 testing days). Gap 2 comprised training data from 2011-01-18 to 2011-10-04 and testing data from 2012-04-02 to 2012-07-19 (51 testing days). Gap 3 comprised training data from 2012-04-02 to 2012-10-12 and testing data from 2013-01-26 to 2013-07-10 (37 testing days). Across individual gaps, the same trends showed were displayed, with the MRNN Big w/SP decoder always achieving the best performance (p < 0.01, signed-rank test with every other decoder, all gaps, except in Gap 2 when comparing to MRNN Small w/SP, p = 0.1, and Gap 3 where it on average achieved a lower $r^2$ than MRNN Small w/SP and MRNN Big no SP).

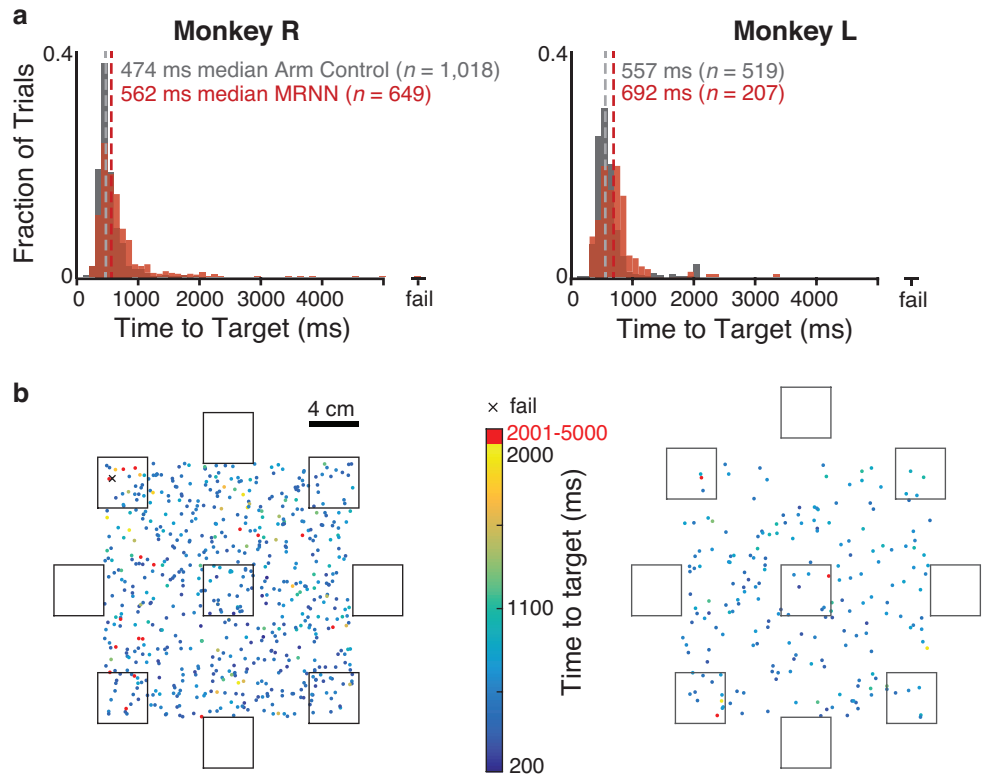

**Supplementary Figure 7. Closed-loop MRNN decoder performance on the Random Target Task.**

Both monkeys were able to use the MRNN decoder to acquire targets across a broad workspace in which targets often appeared at locations that differed from the target locations dominating the training datasets.

(**a**) Histograms of Random Target Task times to target (time of final target entry minus target onset time, not including the 500 ms target hold period) using the MRNN decoder are shown in red. For comparison, histograms of performance on the same task using arm control are shown in gray.

(**b**) Task workspace plots showing the location of each Random Target Task trial's target during MRNN decoder evaluation. Each point corresponds to the center of one trial's target, and its color represents the time it took the monkey to acquire this target. The location of the one failed trial (for monkey R) is shown with a black 'x'. The acquisition area boundaries of the nine Radial 8 Task targets used for the majority of the training data are shown as black squares. Monkey R's data are aggregated across the two experimental sessions in which he performed this task. Monkey L's data are from one session.

**Supplementary Video 1. Example closed-loop performance of the MRNN**
The MRNN was trained using reaching data from 125 recording sessions up through the previous day. The video shows a continuous 90 seconds of monkey R using this decoder to perform the Radial 8 Task. He controls the white cursor and acquires the green target (which turns blue during the 500 ms target hold period). Dataset R.2014.04.03. This is a portion of the data used to generate the drop 0 electrodes condition of Fig. 3.

**Supplementary Video 2. Side-by-side comparison of the MRNN and FIT Sameday decoders after electrode dropping**
During the experiment, the two decoders were evaluated in alternating blocks after the same 3 most important electrodes were dropped. Here we show a continuous 60 seconds of each decoder's closed-loop performance from consecutive blocks. The MRNN (right side) was trained using reaching data from 125 recording sessions up through the previous day, while the FIT Kalman filter (left side) was trained using reaching data from earlier that same day. Dataset monkey R.2014.04.03. This is a portion of the data used to generate the drop 3 electrodes condition of Fig. 3.

**Supplementary Data 1. Array recording quality measurements across this study's recording days**

For task consistency, these analyses were restricted to those recording days when Baseline Block data was collected. Reach direction tuning was calculated as in [32]: for each recording day, we calculated each electrode's average firing rate over the course of each trial (analysis epoch: 200 to 600 ms after target onset) to yield a single data point per trial, and then grouped trials by target location. 69.0 ± 7.7 of monkey R's electrodes (mean ± s.d. across recording days) and 66.3 ± 12.3 of monkey L's electrodes exhibited significantly different firing rates when reaching to at least one of the eight different targets ($p < 0.01$, one-way ANOVA). These tuned electrodes' modulation range, defined here as the trial-averaged rates firing rate difference between reaches to the two targets evoking the highest and lowest rates, was 26.4 ± 5.9 Hz in monkey R monkey (mean ± s.d., averaged first across all electrode pairs in a given recording day, and then over days) and 23.8 ± 5.6 Hz in monkey L. We did not observe cross-talk between electrodes' threshold crossings, consistent with recording spiking activity from electrodes at least 400 μm apart: pairwise electrode cross-correlations, computed using the time-series of firing rates in consecutive non-overlapping 20 ms bins spanning a given day's Baseline Block, was 0.0089 ± 0.0021 in monkey R (mean ± s.d., averaged first across all electrode pairs in a given recording day, and then over days) and 0.0150 ± 0.0058 in monkey L.